\documentclass[reprint,twocolumn,superscriptaddress,secnumarabic,amssymb,nobibnotes,aps,pra]{revtex4-1}

\usepackage{amsmath}    
\usepackage{graphicx}    
\usepackage{verbatim}    
\usepackage{color}           
\usepackage{subfigure}   
\usepackage{hyperref}    
\usepackage{verbatim}  
\definecolor{ao}{rgb}{0.0, 0.5, 0.0}

\begin{document}
\title{
Valence skipping, internal doping and site-selective Mott transition in PbCoO$_3$ under pressure
}

\author{Atsushi Hariki} 
\thanks{These authors contributed equally to this work.}
\affiliation{Department of Physics and Electronics, Graduate School of Engineering, Osaka Prefecture University 1-1 Gakuen-cho, Nakaku, Sakai, Osaka 599-8531, Japan}
\affiliation{Institute for Solid State Physics, TU Wien, 1040 Vienna, Austria}
\author{Kyo-Hoon Ahn} 
\thanks{These authors contributed equally to this work.} 
\affiliation{Institute for Solid State Physics, TU Wien, 1040 Vienna, Austria}
\affiliation{Institute of Physics of the CAS, Cukrovarnick\'{a} 10, 162 00 Praha 6, Czechia}
\author{Jan Kune\v{s}} 
\thanks{kunes@ifp.tuwien.ac.at}
\affiliation{Institute for Solid State Physics, TU Wien, 1040 Vienna, Austria}

\date{\today}

\begin{abstract}
We present a computational study of 
PbCoO$_3$ at ambient and elevated pressure. We employ 
the static and dynamic treatment of local correlation in form
of density functional theory + $U$ (DFT+$U$) and  + dynamical mean-field theory (DFT+DMFT). 
Our results capture the experimentally observed crystal structures and identify the
unsaturated Pb $6s$ -- O $2p$ bonds as the driving force beyond the complex physics of PbCoO$_3$. We provide a geometrical analysis of the structural distortions and we discuss their implications, in particular, the internal doping, which triggers a transition between
phases with and without local moments and a site selective Mott transition in the low-pressure phase.

\end{abstract}

\maketitle

\section{\label{sec:int} Introduction}

Transition metal oxides with perovskite structure host a plethora of interesting electronic phenomena including high-$T_c$ superconductivity, colossal magnetoresistance, metal-insulator transitions, ferro- as well as antiferromagnetism, and charge or orbital ordering~\cite{imada98,khomskii14,Keimer15,Dagotto01}. 
One of the key parameters affecting their physical properties is
filling of the $3d$ orbitals. It can be altered by doping, in the ABO$_3$ perovskite structure usually achieved  by a combination of divalent alkali-earth metals and trivalent rare-earth ions on the A sublattice.
The A-site ions do not participate in formation of the low-energy valence 
and conduction states, playing a passive role of electron donor, 
with rare exceptions such as the Pr ion in the (Pr$_{1-y}$Y$_y$)$_x$Ca$_{1-x}$CoO$_3$~\cite{tsubouchi2002}.
Thanks to quasi-degeneracy of Pr$^{3+}$ and Pr$^{4+}$ valence states
their abundance varies with temperature or pressure~\cite{hejtmanek2010,hejtmanek2013}, leading to internal doping with a substantial impact on the physical properties~\cite{kunes2014b}.

Another path to internal doping is by growing materials with 
valence skipping ions Bi or Pb on the A site. 
A famous example of BiNiO$_3$ exhibits a colossal negative thermal expansion as well as temperature and pressure induced metal-insulator transition~\cite{Ishiwata02,Azuma11,Paul19,Naka16}. Unlike the ${\text{Pr}^{3+}\leftrightarrow\text{Pr}^{4+}}$
crossover, which is an intra-atomic effect, the valence skipping between formal 6$s^0$  (Pb$^{4+}$, Bi$^{5+}$) and 6$s^2$ states (Pb$^{2+}$, Bi$^{3+}$) is a 
metaphor since its origin lies in strongly covalent character of $A$--O bond.
In the Pb$M$O$_3$ ($M$ is a $3d$ metal) series the formal Pb valence changes 
from  Pb$^{2+}$Ti$^{4+}$O$_3$ to Pb$^{4+}$Ni$^{2+}$O$_3$ with a peculiar
behavior for $M$ = Co and Fe in the middle~\cite{Sakai17,Liu20,Belik05,Yu15,Inaguma11,Ye21,Yu15,Wang15}.

In this article, we present a theoretical study of the electronic and crystal structure of PbCoO$_3$ (PCO) and its evolution under pressure. 
Using the density functional theory (DFT) combined with a static and dynamical mean-field treatment
of the electron-electron interaction within the Co $3d$ shell
we show that the physics of PCO arises from the coexistence of several phenomena: (i) valence skipping and internal doping similar to BiNiO$_3$,
(ii) site disproportionation (SD) of the Co sublattice and breathing mode (BM) distortion of CoO$_6$ octahedra as in rare-earth nickelates RNiO$_3$~\cite{Catalano18,Alonso01}
(iii) spin-state crossover as in La$_{1-x}$Sr$_x$CoO$_3$ 
and iv) lattice distortion as in CaCu$_3$Fe$_4$O$_{12}$~\cite{Yamada08,Shimakawa15} associated with tilting
of rigid CoO$_6$ octahedra common to many transition metal perovskites. A combination of
these phenomena leads to complex changes of electronic and structural properties 
with applied pressure, the root cause of which can be traced back to the general tendency of materials to reduce or eliminate their Fermi surface.


At ambient pressure PCO forms a quadruple perovskite structure,
with two distinct Pb sites.
Based on x-ray absorption, photoemission and magnetic susceptibility measurements with a DFT+$U$ calculation, the authors in Ref.~\onlinecite{Sakai17} suggested SD of the Pb$^{2+}$Pb$^{4+}_3$Co$^{2+}_2$Co$^{3+}_2$O$_{12}$ type, which, in addition to the charge gap, survives well above the magnetic transitions temperature of 7.8~K.  Curie-Weiss susceptibility observed at high temperatures was attributed to a high-spin  configuration of the Co$^{2+}$ ion.
Recently, Liu {\it et al}.~\cite{Liu20} reported a pressure-induced insulator-insulator transitions from cubic to tetragonal structure with an intermediate regime between 15 and 30 GPa, in which a bad metal behavior was observed at elevated temperatures.


We employ two approaches. The computationally cheap DFT+$U$ (static mean-field approximation) allows us to find equilibrium atomic positions and lattice
parameters, while we have to admit an unrealistic (ferromagnetic) magnetic order
to capture the high-spin state of Co. We use computationally expensive DFT+ dynamical mean-field theory (DMFT) to capture the fluctuating local moment and dynamical correlation effects
to verify the DFT+$U$ results in the low- and high-pressure structures.


\section{\label{sec:met}Computational Method}
The internal parameters as well as the $c/a$ ratio were optimized for each structure
with the DFT+$U$ method using
the Vienna $ab$ initio simulation package ({\sc vasp})~\cite{Kresse96a,Kresse96b}.
We have employed the local density approximation~\cite{ceperley80,perdew1992} to the exchange correlation potential and the so-called fully localized limit~\cite{Czyifmmode94} form of the double-counting correction in the DFT+$U$ calculations with $U_{\text{eff}}=5.9$~eV.
The atomic positions for all structures can be found in the supplemental material (SM)~\cite{sm}.
The total energies were then calculated for the optimized structures with the all-electron
linearized augmented plane wave method as implemented in {\sc wien2k} code~\cite{blaha20};~see Fig.~\ref{fig:EvV}. 

DFT+DMFT calculations~\footnote{All calculations were performed at the temperature of 300~K.} proceed in two steps:~(i) construction of a tight-binding model from DFT bands for a structure and (ii) solution of the DMFT self-consistent equation~\cite{georges96,kotliar06}. 
In step (i) we use the optimized crystal structures described above, calculate the electronic band structure~\cite{blaha20} with the local density approximation, and construct a lattice model on the basis of Wannier orbitals using {\sc wannier90} code~\cite{wannier90,wien2wannier}. The Wannier projection provides a freedom in choosing the covered orbital space. We use two settings. The $spd$ model spanned by the Co 3$d$, O 2$p$, and Pb $6s$ basis, which represents exactly
the bands in the range -10 -- 2~eV, is used in the DMFT calculations presented in the text.
The $sd$ model spanned
by the  Co 3$d$ and Pb $6s$ basis, represents the Co $3d$-like and Pb $6s$ -- O $2p$ anti-bonding bands.
It was obtained using the band disentanglement technique~\cite{souza2001}, which is used to isolate the Pb $6s$ -- O $2p$ anti-bonding band. The DMFT results
obtained with the $sd$ model (consistent with those obtained with the $spd$ model) are presented in the SM~\cite{sm}.

In step (ii) the tight-binding lattice model is augmented with an on-site interaction term within the Co $3d$
parametrized in terms of the Coulomb $U$ and Hund's $J$ parameters~\cite{pavarini1,pavarini2}.
Following previous studies on LaCoO$_3$~\cite{krapek12,Karolak15}, we use ($U$, $J$) = (6.0~eV, 0.8~eV) for the $spd$ model and ($U$, $J$) = (3.0~eV, 0.6~eV) for the $sd$ model~\cite{sm}. The fully localized form of the double-counting correction is employed~\cite{karolak10,Karolak15,Czyifmmode94,Park14}. The strong-coupling continuous-time quantum Monte-Carlo method~\cite{werner06,boehnke11,hafermann12,hariki15} is used to solve the auxiliary Anderson impurity model (AIM) in the DMFT calculation. The spectral functions are calculated from the self-energy analytically continued by the maximum entropy method~\cite{jarrell96}. Weights of the atomic states given as the partial traces
of the reduced density matrix and the imaginary time local spin-spin correlation function 
$\langle S_z(\tau)S_z(0)\rangle$ are measured directly in the quantum Monte-Carlo simulation.


\section{\label{sec:res}Results and discussion}
\begin{figure}[t] 
\includegraphics[width=1.0\columnwidth]{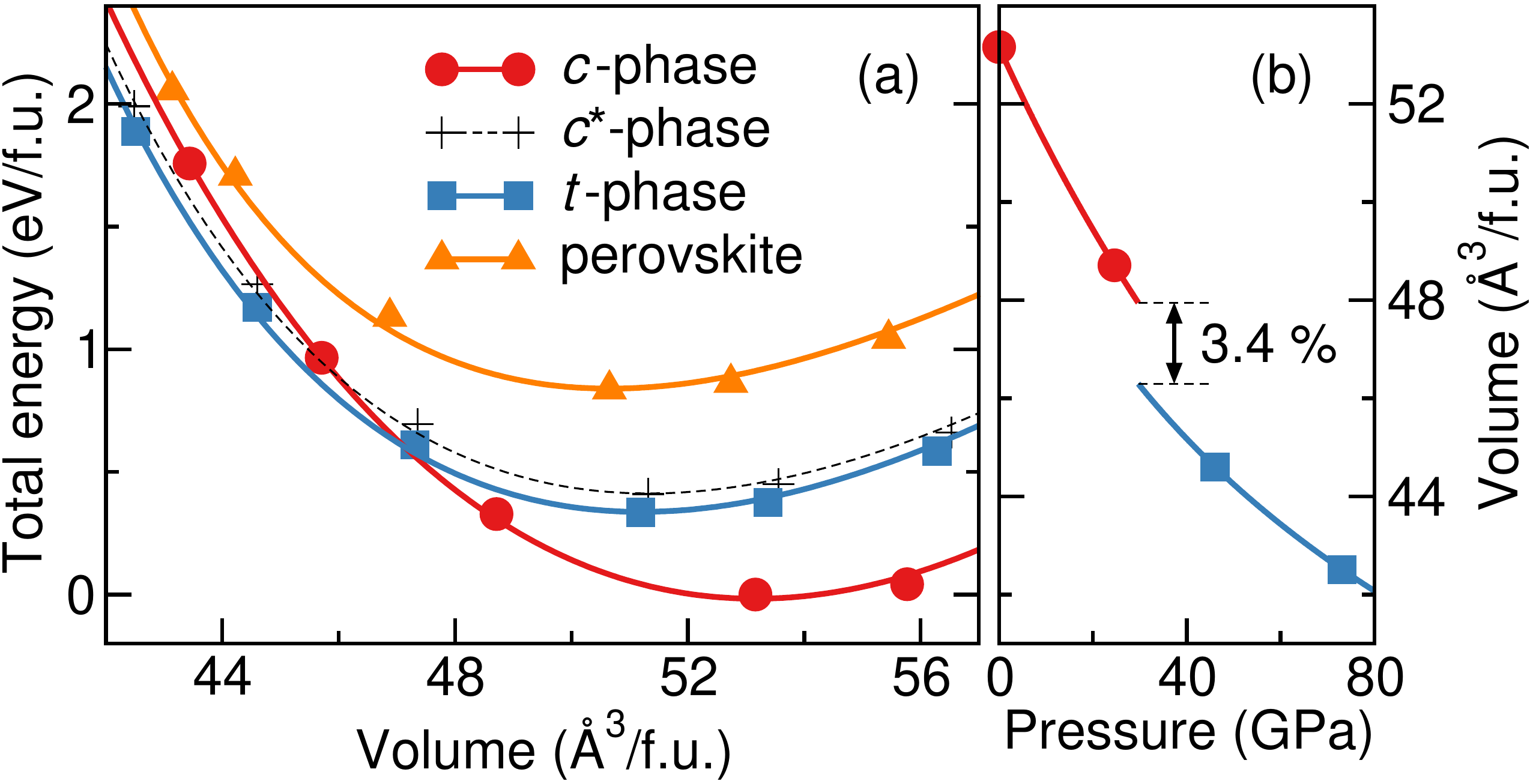}
\caption{(a) Energy vs volume curves for various structures obtained with DFT+$U$ of $U_{\rm eff}=5.9$ eV. The bottom of the $c$-phase is set to zero energy. (b) Volume vs pressure curves with $P_c=30$ GPa, obtained by common tangent construction from (a).}
\label{fig:EvV}
\end{figure}

In Fig.~\ref{fig:EvV} we compare the energy vs volume dependencies 
for several structures: simple perovskite ($Pm\bar{3}m$), cubic quadruple perovskite with BM distortion ($c$-structure; $Pn\bar{3}$), cubic quadruple perovskite without BM distortion ($c^*$-structure; $Im\bar{3}$) and tetragonal ($t$-structure; $I4/mmm$)
obtained with DFT+$U$ approach. The calculations capture the experimental observation of the $c$-structure having the lowest energy at ambient pressure and the $t$-structure having the lowest energy at elevated pressure. The transition pressure of 30~GPa as well as the discontinuous volume change of 3.4\% agrees fairly well with the experimental observations~\cite{Liu20}. 
We did not identify the intermediate pressure phase of Ref.~\onlinecite{Liu20}, however, a near degeneracy of several phases in the vicinity of the transition indicates a flat energy landscape with the possibility of additional stable structures possibly with larger unit cell.

\subsection{Parent perovskite structure}

\begin{figure}[t] 
\includegraphics[width=1.0\columnwidth]{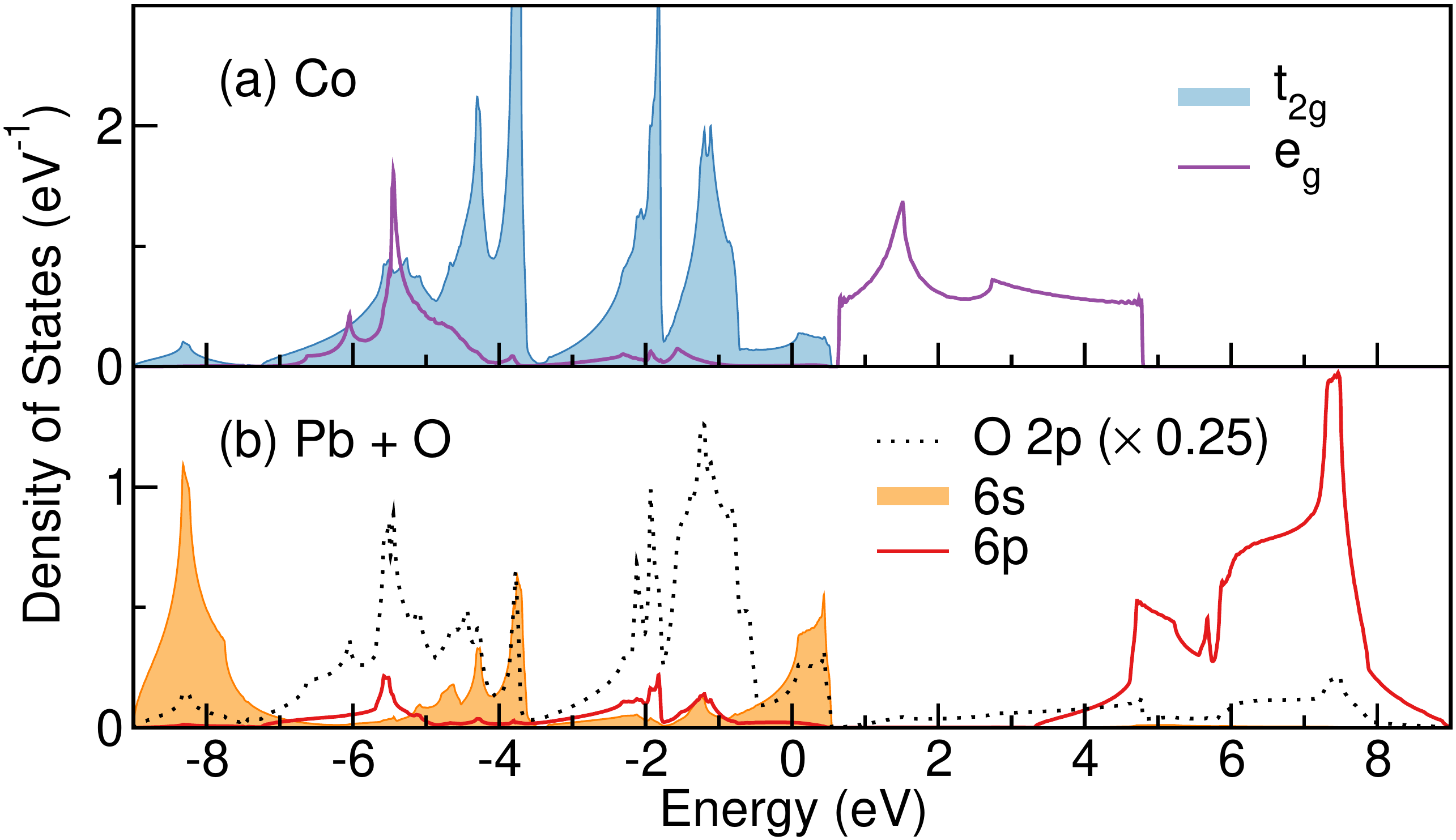}
\caption{Orbital-resolved densities of states of simple perovskite structure obtained with DFT+$U$ of $U_{\rm eff}=5.9$ eV. The orbitals are defined as atomic sphere projections of given angular symmetry.}
\label{fig:perovskite}
\end{figure}

While the simple perovskite structure is unstable for all studied volumes shown in Fig.~\ref{fig:EvV} it provides a useful reference and sets the energy scales in PCO. 
The DFT+$U$ spectra in Fig.~\ref{fig:perovskite} reveal that
the main effect of electron-electron interaction within the Co $3d$ shell is to enhance the octahedral crystal field and open a gap between Co $t_{2g}$ and $e_g$ states. 
A strong Pb $6s$ -- O $2p$ hybridization positions the half-filled anti-bonding band at the Fermi level. Moderate Pb $6p$ -- O $2p$ hybridization distinguishes PCO from the ferroelectric perovskite BiCoO$_3$, where strongly hybridized Bi $6p$ -- O $2p$ bonding states appear in the vicinity of the Fermi level while the Bi $6s$ -- O $2p$ hybridization is substantially weaker than in PCO. This difference results from deeper Bi $6s$ and $6p$ levels;~see the SM~\cite{sm}. The combination of $6s$ -- O $2p$ anti-bonding and $6p$ -- O $2p$ bonding bands is decisive for the occurrence and type of structural distortions in Pb and Bi oxides~\cite{Watson99}. We conjecture the dominance of Pb $6s$ -- O $2p$ hybridization to be the reason for approximately centrosymmetric coordination of Pb sites in all PCO structures~\cite{Liu20}.

Unlike in rare-earth cobaltites, where the tilting of CoO$_6$ is driven by electrostatic instability of the small rare-earth ion in the Co--O matrix, the structural distortions in PCO and other active $A$-site perovskites~\cite{Paul19,Paul21} originate from covalent Pb--O bonding. Formulated in the language of electronic bands, a structural distortion leads to energy lowering if it opens a (partial) gap at the Fermi level while keeping the elastic energy cost low. A textbook example is the Peierls instability of one-dimensional (1D) systems driven by perfect Fermi surface nesting. In general 3D systems the Fermi surface nesting can only be approximate and a finite distortion may be necessary to achieve energy lowering. In fact, the numerical results below
classify the distortion-induced band splitting rather as a strong coupling effect (splitting is larger or comparable to the bandwidth) and thus Fermi surface nesting does not play an important role.

While the calculations presented in Fig.~\ref{fig:EvV} take the whole system into account, understanding them requires breaking the system into functional components and identification
of leading interactions within and between them. Such simplifications can usually be justified 
only {\it a posteriori}. We start our discussion by analyzing the geometry of the distorted structures.

\begin{figure}[t]
\includegraphics[width=1.0\columnwidth]{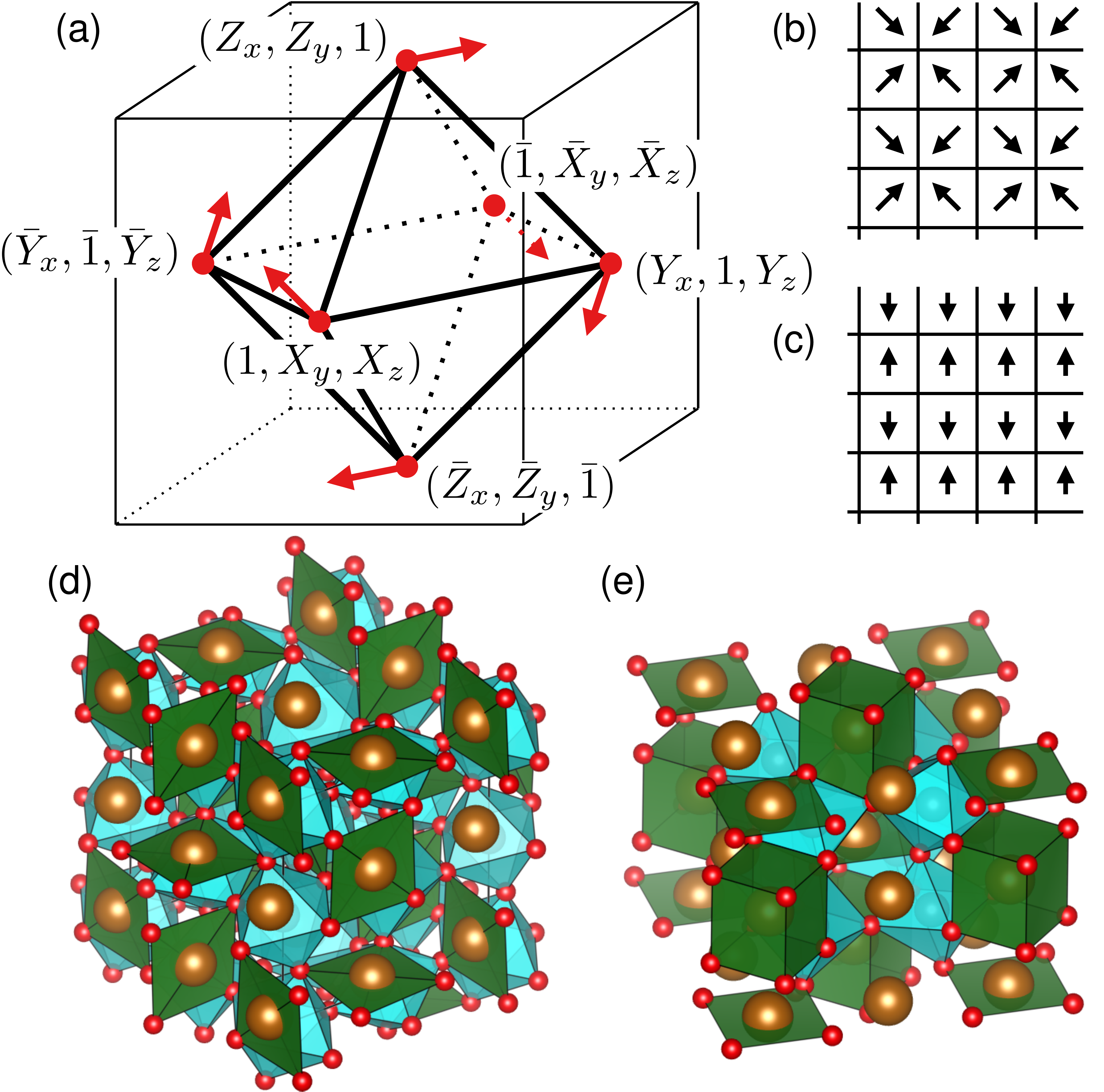}
\caption{(a) O-displacements due to tilting of the CoO$_6$ octahedron. (Red dots mark the O ions, Pb ions sit in the corners of the cube.) Diagonal (b)
and parallel (c) displacements of O ions in the PbO layer. (Pb ions sit in the vertices, O ions in the square centers.) The $c$ (d) and $t$ (e) structures
with O sites (red) and Pb sites (gold). The green plaquettes and cuboids belong to Pb sites with a saturated Pb $6s$-O 2$p$ bond.}
\label{fig:tilt}
\end{figure}

\subsection{Geometry considerations}

Distortions of simple perovskite structure into the low-pressure cubic ($c$) and high-pressure tetragonal ($t$) structure can be viewed as tilts of (approximately) rigid CoO$_6$ octahedra,
shown in Fig.~\ref{fig:tilt}(a). Small tilting of a CoO$_6$ octahedron corresponds to shifting of the O atoms within the PbO planes subject to constraints $Z_x=-X_z$, $Z_y=-Y_z$, and $X_y=-Y_x$, which reduce the number of independent parameters to 3 as expected for a 3D rotation of a rigid body. Sharing corners by neighboring octahedra implies further constraints on possible O-displacements in the PbO and CoO$_2$ planes. Let us consider layers parallel to the $(0,0,1)$ plane.
Allowed O-displacements in the PbO layer fulfill  $Z_x(\mathbf{R}+\mathbf{e}_x)=-Z_x(\mathbf{R})$ and 
$Z_y(\mathbf{R}+\mathbf{e}_y)=-Z_y(\mathbf{R})$~\footnote{Here $\mathbf{R}$ is a position vector on the cubic perovskite lattice and $\mathbf{e}_x$ and $\mathbf{e}_y$ denote lattice basis vectors.}, which reduces the 
the number of independent parameters on the $N\times N$ lattice from $2N^2$ for the
case of independent octahedra to $2N$. 
To extend this reasoning to the 3D structure we observe the following: (i) The displacement pattern in one PbO layer uniquely determines the displacements in all remaining PbO layers. 
(ii) After fixing $Z_x(\mathbf{R})$ and $Z_y(\mathbf{R})$, it remains to determine 
the $X_y(\mathbf{R})$ in the adjacent CoO$_2$ layer. Fixing this parameter for an arbitrary
O site, e.g., $X_y(0,0,R_z)$, uniquely determines the displacements for all remaining O sites in the layers and thus is a sole free parameter.
(iii) $X_y(0,0,R_z)$ in distinct CoO$_2$ layers are independent. As a result the allowed tilts on the $N\times N \times N$ lattice are described by $3N$ independent parameters (no periodicity assumed). The allowed tilts are severely limited if the inversion symmetry of the Pb sites is to be preserved.
In the $(0,0,1)$ plane it adds the conditions 
$Z_y(\mathbf{R}+\mathbf{e}_x)=Z_y(\mathbf{R})$ and $Z_x(\mathbf{R}+\mathbf{e}_y)=Z_x(\mathbf{R})$ to those above, which implies that
the displacement of a single O atom determines uniquely the whole layer. Extending
the argument to 3D shows that a tilt of a single CoO$_6$ octahedron (three parameters) determines the pattern throughout the entire crystal.

Symmetry suggests to search for local energy minima for displacements along high symmetry directions, i.e. along the cubic edges or face diagonals as shown in Figs.~\ref{fig:tilt}(b) and \ref{fig:tilt}(c). Both the $c$ and $t$ structures can be obtained from 
these two patterns. The $c$-structure, shown in Fig.~\ref{fig:tilt}(d), arises from
diagonal displacements in all three PbO planes (cubic faces). As a results
$3/4$ of the Pb atoms belong to PbO$_4$ plaquettes with short Pb--O bonds 
and $1/4$ of Pb atoms remain ``lone". 
The $t$-structure, shown in Fig.~\ref{fig:tilt}e, originates from diagonal 
displacement in the $(0,0,1)$ plane and $z$-parallel displacement in 
the $(1,0,0)$ and $(0,1,0)$ planes. This gives rise to three inequivalent Pb sites:
$1/4$ of Pb atoms forms PbO$_4$ plaquettes as in the $c$-structure, $1/4$ of Pb
atoms form short-bond cuboids 
and $1/2$ of Pb atoms remain lone.

\begin{figure}[t] 
\includegraphics[width=0.98\columnwidth]{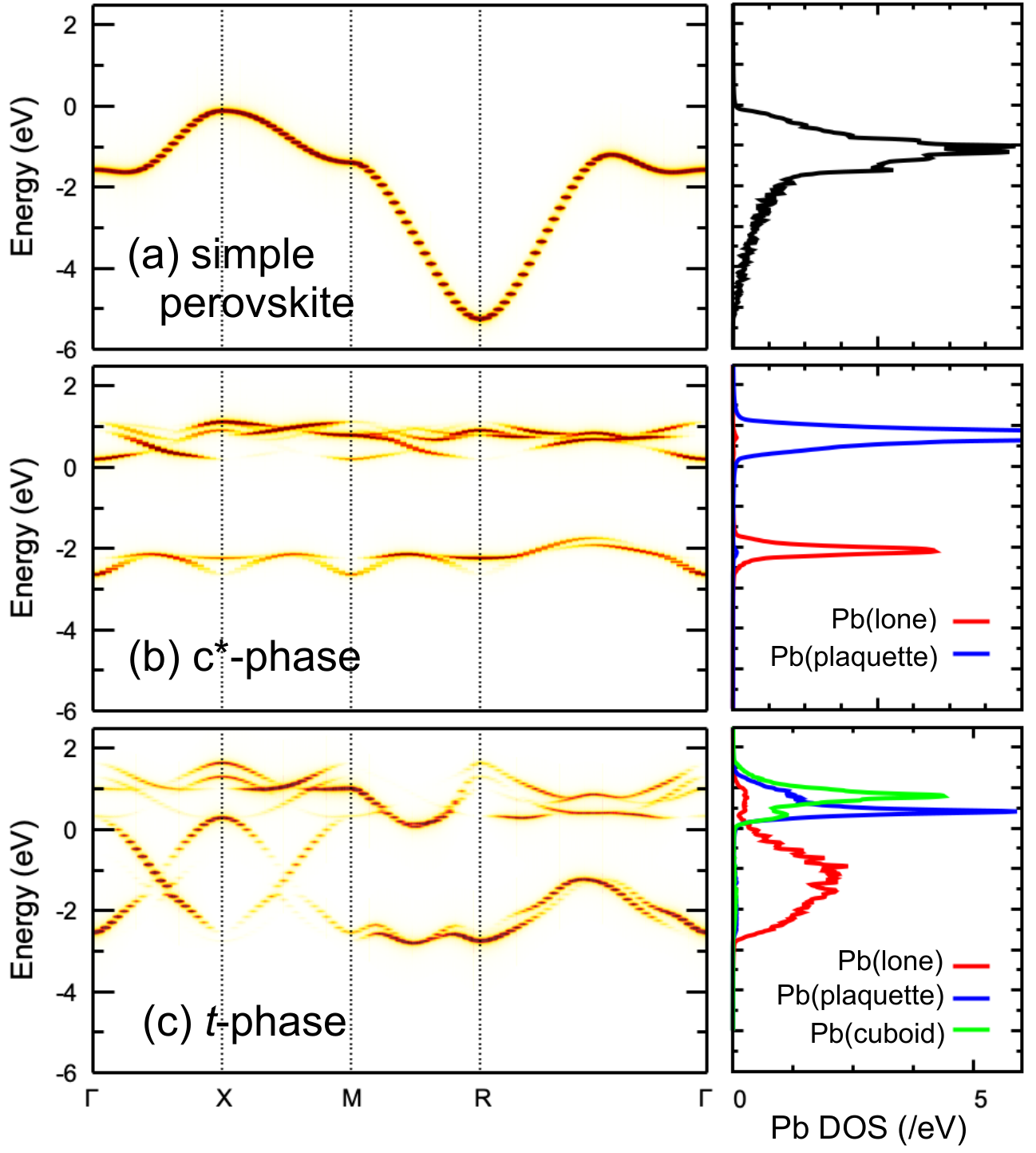}
\caption{Dispersion of isolated Pb 6$s$ anti-bonding state in (a) simple perovskite, (b) $c^*$-phase, and (c) $t$-phase. All bands are unfolded into the Brillouin zone of the simple perovskite~\cite{ku2010}. The right panels show the densities of states projected on inequivalent Pb sites. Note that $E=0$ does not mean Fermi energy in this construction.}
\label{fig:unfold}
\end{figure}

\subsection{Low-pressure structure}
The total energies in Fig.~\ref{fig:EvV} show that all the considered structural distortions lead to energy lowering. The absence of the Co $3d$ spectral density in the vicinity of the Fermi level, shown in Fig.~\ref{fig:perovskite}, as well as the fact that different distortion patterns ($t$ and $c^*$) lead to similar energy lowering suggest that the distortion is driven by Pb--O bonding. 
In the reciprocal space, this can be understood as a tendency to open a gap (at the Fermi level) in the Pb--O anti--bonding band, similar to the Peierls instability. In the direct space, this can be viewed as disproportionation of Pb--O bonds into saturated ones (empty anti-bonding state) and non-bonds associated with lone Pb ions (occupied anti-bonding state). 

The $t$-distortion splits the anti--bonding band into a low-energy doublet 
and two singlets. The $c^*$-distortion splits the anti--bonding band into a low-energy singlet and a triplet, referring to a Pb$_4$Co$_4$O$_{12}$ unit cell in both cases. 
To demonstrate this behavior we have constructed a $ds$ model spanning the Co $3d$ and anti-bonding Pb $6s$ -- O $2p$ bands, where the O $2p$ are only implicitly included in the Co and Pb centered Wannier orbitals.
This construction is justified by the strong Pb -- O, which well separates the \textcolor{black}{bonding} and anti-bonding bands;~the exception is near the $R$ point of the Brillouin zone, where Pb -- O hybridization vanishes, but the corresponding states play no role in the studied  physics as they lie far below the Fermi level.
To further reduce complexity of the problem, we switch off the $d$--$s$ hybridization (its effect is shown in the SM~\cite{sm}) and thus isolate the physics of the Pb -- O anti-bonding band. In Fig.~\ref{fig:unfold} we
show the corresponding band structures unfolded~\cite{ku2010} to the Brillouin zone of the simple perovskite structure.

\begin{figure*}[t] 
\includegraphics[width=2.0\columnwidth]{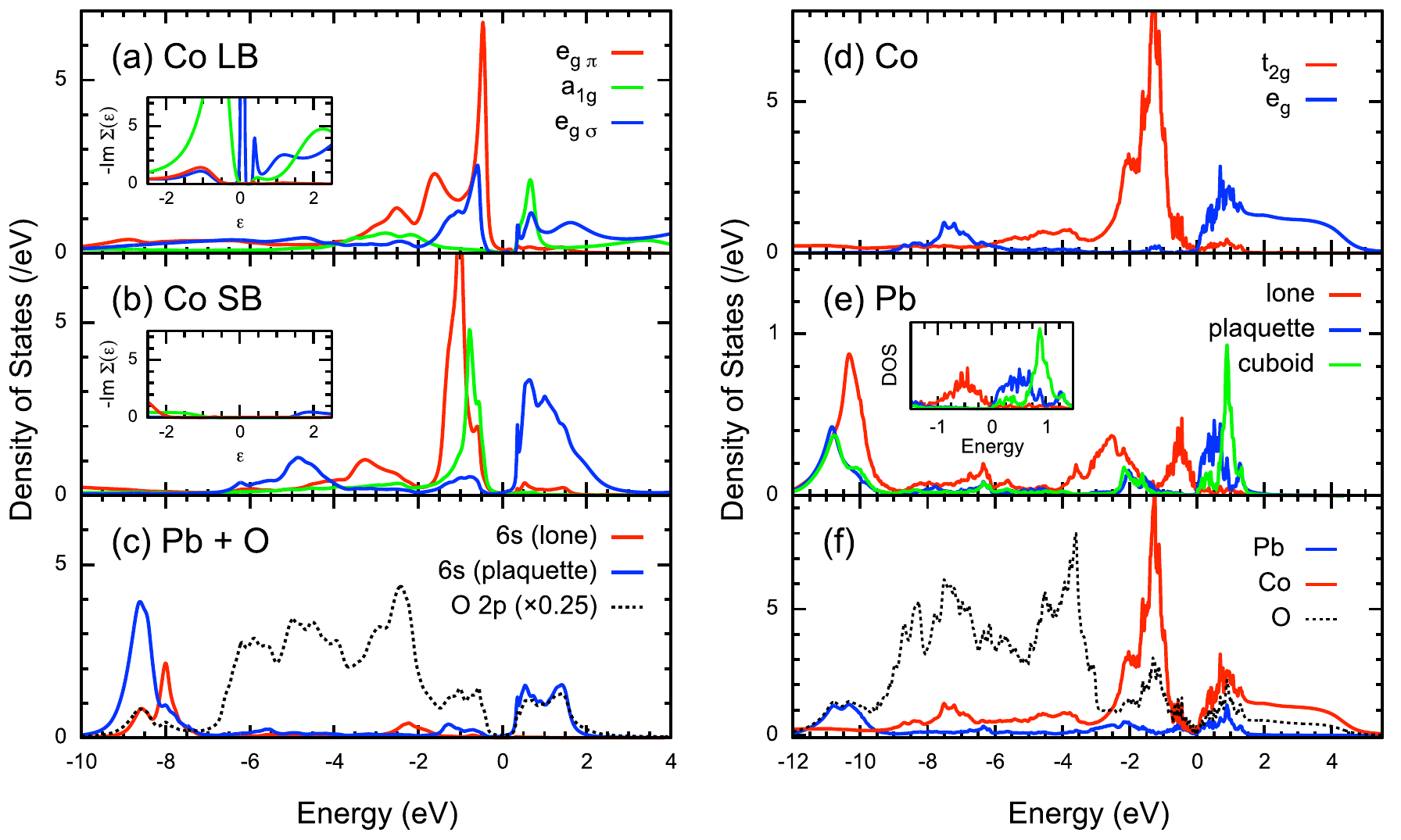}
\caption{Left:~DFT+DMFT orbital-resolved densities of states
in the $c$ structure (low-pressure structure)
on (a) Co LB site, (b) Co SB site, and (c) Pb lone and \textcolor{black}{plaquette}, and O sites. (The orbitals correspond to the Wannier basis.) 
The imaginary parts of the DMFT self-energies $\Sigma(\omega)$ of the Co 3$d$ orbitals are shown in the insets of panels (a) and (b).
Right:~DFT+DMFT orbital-resolved projected density of states in the $t$ structure (high-pressure structure) on (d) Co site, 
\textcolor{black}{(e) Pb lone, plaquette, and cuboid.}
The element resolved spectra are shown in panel (f). The inset in panel (e) shows the low-energy region.
}
\label{fig:ldadmft_dos}
\end{figure*}

While the $c^*$-distortion already lowers the energy relative to the simple perovskite,
it leaves the Pb--O bonds of the triplet unsaturated -- the anti-bonding state is occupied by two electrons. Further energy lowering is possible by transfer of the excess electrons from the triplet Pb--O anti-bonding state to the Co bands. In a non-interacting system, such a transfer cannot lead to energy lowering, because the extra electron must go to the unoccupied Co $e_g$ bands, located above the Pb--O anti-bonding state. In an interacting system, however, doping is not equal to simply filling or emptying of electronic bands
as is known for hole-~\cite{Podlesnyak2008,sboychakov2009,Augustinsky2013} as well as electron-~\cite{tomiyasu2017} doped cobaltites.
Sakai {\it et al.}~\cite{Sakai17} showed that the 
$c^*$-distortion is accompanied by transfer of, on average, $1/2$ electron per Co ion. This state is
stabilized by site disproportionation of Co into high-spin Co$^{2+}$ and low-spin Co$^{3+}$. 

\begin{figure}[t] 
\includegraphics[width=0.98\columnwidth]{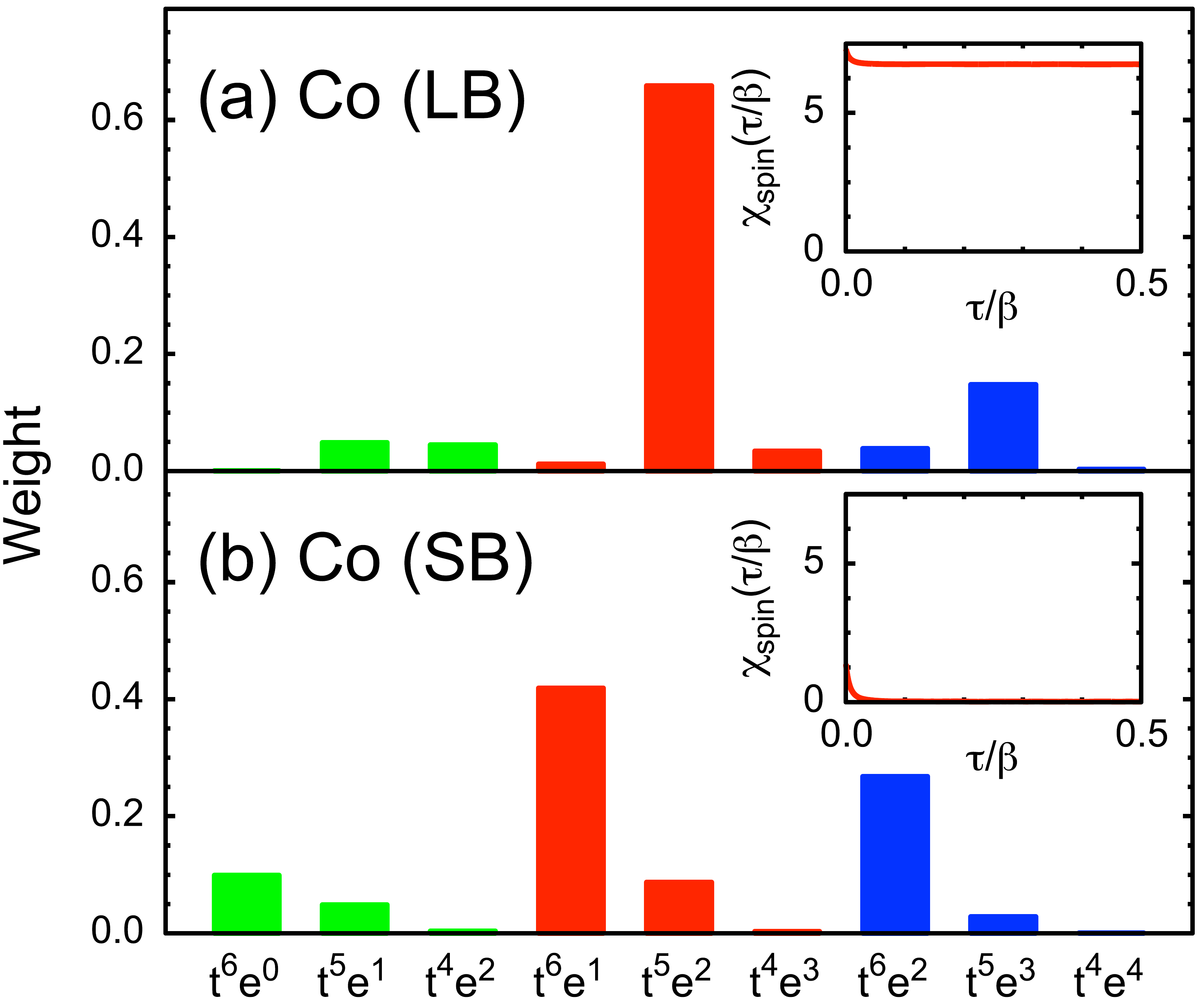}
\caption{Weights of dominant atomic states in $d^6$ (green), $d^7$ (red) and $d^8$ (blue) occupation sectors at (a) Co LB site and (b) Co SB site calculated by the DFT+DMFT method. The spin-correlation function $\chi_{\rm spin}(\tau)$ calculated at the Co LB and the Co SB site is shown in the inset of panel (a) and (b), respectively.}
\label{fig:c_weight}
\end{figure}

While our total energy calculations assumed ordered magnetic moments on the Co sites, because only such a state can be described by the static DFT+$U$ approach, PCO is a paramagnetic insulator down to 7.8~K~\cite{Sakai17} before it magnetically orders. To capture the paramagnetism and dynamical correlation effects as well as to allow a ``fair" competition between disproportionated insulating and uniform metallic states~\footnote{Static approaches such as DFT+$U$ are known to exaggerate ordering tendencies.}, we have performed DFT+DMFT calculation in the $c$-structure using the structural parameters available for ambient pressure~\cite{Sakai17,sm}. The results, summarized in Figs.~\ref{fig:ldadmft_dos} and \ref{fig:c_weight}, confirm the formation of a local moment on $1/2$ of Co sites. It leads
to partial filling of the Co $e_g$--O $2p$ anti-bonding states and weakens the Co--O bond - a physicist's way of saying that high-spin state has a larger ionic radius.
\textcolor{black}{
This results in an expansion (2.059~\AA) of the Co--O bonds with high-spin Co, the Co long-bond (LB) site, 
relative to the bonds (1.925~\AA)
with low-spin Co, the Co short-bond (SB) site.} 
Overall the $c$-structure expands with respect to the competing structures with no high-spin Co,
which causes its demise at higher pressure.

Origins of the charge gaps on Co SB and LB sites are quite different. The SB site resembles a band insulator with a crystal-field gap between the occupied $t_{2g}$ and empty $e_g$ states of Co$^{3+}$ ion. The $e_g$ orbitals participate
in the Co--O $\sigma$-bonds, which is reflected in the $d^6\leftrightarrow d^7$ charge fluctuations shown
in Fig.~\ref{fig:c_weight}b. 
On the LB site both the $t_{2g}$ (the $a_{1g}$ to be precise) and $e_g$ states are partially occupied. The gap is an effect of strong correlations, 
which are on one-particle level manifested by peaks in the self-energy at, or close to, the Fermi level. Carrying this characteristic of the Mott insulator, 
the LB site resembles an isolated Co$^{2+}$ ion in the high-spin state. The fluctuating local moment is reflected in spin-spin correlations $\chi_{\text{spin}}(\tau)=\langle S_z(\tau)S_z(0)\rangle$, shown in the inset of Fig.~\ref{fig:c_weight}(a) and the suppressed charge fluctuations, Fig.~\ref{fig:c_weight}a.
The charge disproportionation is rather small ($n^{\rm LB}_d-n^{\rm SB}_d\approx 0.1$) and one can observe a difference in charge and spin dynamics rather than in average occupations, similar to rare-earth nickelates~\cite{Johnston14,Green16,Park12}. 


\subsection{High-pressure structure}
In Figs.~\ref{fig:ldadmft_dos}(d)--(f)
we show the orbital resolved spectral densities in the high-pressure $t$-structure obtained by the DFT+DMFT method for the theoretical structural parameters~\cite{sm}
corresponding to $P=73$~GPa 
of Fig.~\ref{fig:EvV}(b).
The Co bands resemble the simple perovskite spectrum with the occupied $t_{2g}$ band separated by a crystal-field gap from the empty $e_g$ bands. The location of Fermi level inside the crystal-field gap makes the Co $d$ states play rather passive role. 

The $t$-distortion distinguishes three Pb sites: the lone site (Pb1), the plaquette site (Pb2) with a saturated bond to four O neighbors and the cuboid (Pb3) with a saturated bond to eight O neighbors. The Pb-projected spectra 
[Fig.~\ref{fig:ldadmft_dos}(e)]
reveal a pseudogap separating the Pb1 and Pb2+Pb3 anti-bonding bands. The Pb2 and Pb3 peaks above the Fermi level are rather narrow, which reflects the isolated location of both PbO$_4$ plaquettes and PbO$_8$ cuboids in the structure (O ions participating in a saturated bonds are not shared by more than one Pb ion). The Pb1 anti-bonding band exhibits a broader two-peak structure from $-4$ to 0~eV. It originates from a broad Pb1 anti-bonding band, see Fig.~\ref{fig:unfold}(c), which is split by hybridization with a narrow Co $t_{2g}$ band;~see the SM~\cite{sm} for an explicit demonstration. The width of this parent Pb1 band is noticeably larger than its Pb2 and Pb3 counterparts. This is because the lone Pb1 sites within the $xy$ plane, while having longer Pb-O bonds, are interconnected via shared oxygen sites giving
rise to large in-plane dispersion, Fig.~\ref{fig:unfold}(c).

\subsection{Magnetic and transport properties}
Finally, we comment on the magnetic and transport properties of PCO. The calculated spin dynamics at ambient pressure reveals
the double-perovskite-like structure of Co in nominally $S=3/2$ and $S=0$ states observed in the experiment~\cite{Sakai17}. The separation
of the high-spin Co ions by magnetically inactive low-spin ions is consistent with the low ordering temperature. The high-pressure $t$-phase hosts low-spin Co ions and a small gap or \textcolor{black}{pseudogap}. Therefore low susceptibility is expected over the whole temperature range.

Calculation of resistivity with the present approach may miss important effects such as structural disorder or electron-phonon scattering. Therefore, we focus on overall trends, which can be estimated from the spectral functions. The main observations of Liu {\it et al.}~\cite{Liu20} can be summarized as follows:~(i) resistivity in the $c$-phase drops with temperature substantially faster than in $t$-phase, (ii) resistivity in the $c$-phase increases with pressure, (iii) resisitivity in the $t$-phase decreases with pressure, but the trend possibly reverses at higher temperatures~\footnote{The experiments~\cite{Liu20} performed only up to the room temperature found the upturn only at intermediate pressures.}. The observation (i) is consistent with finding a larger gap in the low-pressure 
$c$-phase. The pressure dependence (ii) can be understood assuming that the charge transport is controlled by the crystal-field gap, which increases with pressure. The $t$-phase exhibits a pseudogap between the Pb--O anti-bonding bands, i.e., the gap is smaller than in the $c$-phase, and the character \textcolor{black}{of} the valence and conduction bands is different, which is consistent with (i). 
The decrease of resisitivity with pressure (iii) may simply reflect increasing quasi-particle mobility (band broadening), but naming a specific mechanism based on the present results would be rather speculative. 
The upturn of resistivity (iv) at high temperature is consistent with a small gap or pseudogap situation
in which a thermal excitation of charge carriers competes with increasing electron-electron or electron-photon scattering. Such a behavior is known for example from small-gap semiconductor FeSi~\cite{Madrus1995,Kunes2008}.

\section{\label{sec:con}Conclusions}

We have presented a computational study of the electronic structure of PbCoO$_3$ (PCO) under pressure using DFT+$U$ and DFT+DMFT approaches. Our results correctly capture the low- and high-pressure structures as well as their insulating character and magnetic properties.
As pointed out in the experimental studies~\cite{Sakai17,Liu20} PCO hosts several interesting effects observed separately in other materials: valence skipping, site disproportionation, internal doping and spin state transition. These effects are a consequence of the covalent Pb--O bond, which distinguishes PCO
from ABO$_3$ perovskites with rare-earth or alkali-earth metals on the A site, and correlation effects 
on the Co site.

Analyzing the numerical results we have attempted to disentangle these effects and put PCO into context of other transition metal perovskites. Starting from a hypothetical simple perovskite structure, we find that the key mechanism determining the crystal structure is an instability of the unsaturated Pb $6s$ -- O $2p$ bond. Unlike in other materials such as BiCoO$_3$ or \textcolor{black}{PbVO$_3$}, where A $6s$ -- O $2p$ anti-bonding state is nominally occupied and the A $6p$ play an important role in non-centrosymmetric displacement of the A ion, Pb $6p$ states in PCO do not play any important role. 
The structural distortion thus distinguishes the Pb sites into one with a saturated bond (empty anti-bonding state) and non-bonding ones (occupied anti-bonding state). We have shown that both the
distortion into the low-pressure cubic $c$-structure and the high-pressure tetragonal $t$-structure can be viewed to the leading order as tilts of rigid CoO$_6$ octahedra, which preserve the inversion symmetry of the Pb site. While the $c$-structure allows opening of a larger gap in the Pb--O anti-binding band, a transfer $1/2$ electron to nominally Co $3d$ band is necessary to position the gap at the Fermi level.
This internal charge transfer is stabilized by disproportionation of Co into a high-spin site with Mott insulator like spectral density and low-spin band insulator like spectral density. The price paid is the 
expansion of the O$_6$ octahedron around the high-spin site due to reduced Co $e_g$--O $2p$ covalency.
This results in the so-called breathing distortion and overall expansion of the structure, which makes it less favorable at elevated pressure, and it is eventually replaced by the non-magnetic tetragonal structure.

Internal doping is a clean way to qualitatively modify materials properties. The rich physics of
PCO provides an example of its realization.


\begin{acknowledgments}

We thank M.~Azuma, K.~Shigematsu, T.~Nishikubo, T.~Uozumi and A.~Kauch for valued discussions.
A.H is supported by Grants-in-Aid for Scientific Research on JSPS KAKENHI (Grant No.~21K13884).
K.H.A is supported by the Czech Science Foundation Projects (Grant No.~19-06433S and 18-12761S).
J.K is supported by the European Research Council (ERC) under the European Union's Horizon 2020 research and innovation programme (Grant No.~646807-EXMAG).
Computational resources were supplied by the Vienna Scientific Cluster (VSC)
and the project e-Infrastruktura CZ (e-INFRA LM2018140) provided within the program Projects of Large Research, Development and Innovations Infrastructures.

\end{acknowledgments}



\bibliography{main}

\end{document}